\documentclass[
reprint,
superscriptaddress,
showkeys,
showpacs,
amsmath,amssymb,
aps,
pra,
]{revtex4-1}

\usepackage{graphicx}% Include figure files
\usepackage{sidecap}
\usepackage{dcolumn}% Align table columns on decimal point
\usepackage{upgreek}
\usepackage{mathrsfs}
\usepackage{amsmath}
\usepackage{setspace}%ʹ�ü�������
\usepackage{bm}
\usepackage{gensymb}
\usepackage[breaklinks=true,colorlinks=true,linkcolor=blue,urlcolor=blue,citecolor=blue]{hyperref}
\allowdisplaybreaks[4]
\begin{document}
	\preprint{APS/123-QED}
	
	%\begin{CJK*}{GBK}{song}
	\title{Dissipative time crystal in a strongly interacting Rydberg gas}
	
	\author{Xiaoling Wu}
	\thanks{These authors contributed equally to this work}
	\affiliation{State Key Laboratory of Low Dimensional Quantum Physics, Department of Physics, Tsinghua University, Beijing 100084, China}
	
	\author{Zhuqing Wang}
	\thanks{These authors contributed equally to this work}
	\affiliation{State Key Laboratory of Low Dimensional Quantum Physics, Department of Physics, Tsinghua University, Beijing 100084, China}
	
	\author{Fan Yang}
	\thanks{These authors contributed equally to this work}
	\affiliation{Center for Complex Quantum Systems, Department of Physics and Astronomy, Aarhus University, DK-8000 Aarhus C, Denmark}

	\author{Ruochen Gao}
	\affiliation{State Key Laboratory of Low Dimensional Quantum Physics, Department of Physics, Tsinghua University, Beijing 100084, China}
	
	\author{Chao Liang}
	\affiliation{State Key Laboratory of Low Dimensional Quantum Physics, Department of Physics, Tsinghua University, Beijing 100084, China}
	
	\author{Meng Khoon Tey}
	\affiliation{State Key Laboratory of Low Dimensional Quantum Physics, Department of Physics, Tsinghua University, Beijing 100084, China}
	\affiliation{Frontier Science Center for Quantum Information, Beijing 100084, China}
	\affiliation{Hefei National Laboratory, Hefei, Anhui 230088, China}
	
	\author{Xiangliang Li}
	\email{lixl@baqis.ac.cn}
	\affiliation{Beijing Academy of Quantum Information Sciences, Beijing 100193, China}
	
	\author{Thomas Pohl}
	\email{thomas.pohl@itp.tuwien.ac.at}
	\affiliation{Institute for Theoretical Physics, TU Wien, Wiedner Hauptstraße 8-10/136, A-1040 Vienna, Austria}
	
	\author{Li You}
	\email{lyou@mail.tsinghua.edu.cn}
	\affiliation{State Key Laboratory of Low Dimensional Quantum Physics, Department of Physics, Tsinghua University, Beijing 100084, China}
	\affiliation{Frontier Science Center for Quantum Information, Beijing 100084, China}
	\affiliation{Hefei National Laboratory, Hefei, Anhui 230088, China}
	\affiliation{Beijing Academy of Quantum Information Sciences, Beijing 100193, China}

	\begin{abstract}
		The notion of spontaneous symmetry breaking has been well established to characterize classical and quantum phase transitions of matter, such as in condensation, crystallization or quantum magnetism. Generalizations of this paradigm to the time dimension can lead to a time crystal phase, which spontaneously breaks the time translation symmetry of the system. Whereas the existence of a continuous time crystal at equilibrium has been challenged by no-go theorems, this difficulty can be circumvented by dissipation in an open system. Here, we report the experimental observation of such dissipative time crystalline order in a room-temperature atomic gas, where ground-state atoms are continuously driven to Rydberg states. The emergent time crystal is revealed by persistent oscillations of the photon transmission, and we show that the observed limit cycles arise from the coexistence and competition between distinct Rydberg components. The nondecaying autocorrelation of the oscillation, together with the robustness against temporal noises, indicate the establishment of true long-range temporal order and demonstrates the realization of a continuous time crystal.
	\end{abstract}
	
	\maketitle
	
	The search for emergent many-body phases is among the central objectives of quantum physics \cite{sondhi1997continuous}. While the concept of phase transitions in thermal equilibrium is well developed, the presence of driving and dissipation can result in a rich phenomenology that has no counterpart in equilibrium \cite{eisert2015quantum}, such as ubiquitous self-organization effects in physics, biology, and economics \cite{haken2006information}. In particular, such nonequilibrium processes can facilitate a novel dynamical phase that spontaneously breaks time translation symmetry \cite{kessler2019emergent,buvca2019non,dogra2019dissipation,dreon2022self}, commonly referred to as a time crystal \cite{shapere2012classical,sacha2017time,zhang2017observation,choi2017observation,rovny2018observation,riera2020time,randall2021many,kyprianidis2021observation,kessler2021observation,mi2022time,taheri2022all}. In analogy to crystals in space, a continuous time crystal (CTC) phase has an order parameter with self-sustained oscillations, even though the system is driven in a continuous manner \cite{wilczek2012quantum,nozieres2013time,bruno2013impossibility,watanabe2015absence,iemini2018boundary,bakker2022driven,carollo2022exact,krishna2023measurement,nie2023mode}. Remarkably, the spontaneous time translation symmetry breaking associated with a dissipative CTC has been recently observed with atomic Bose-Einstein condensate in an optical cavity \cite{kongkhambut2022observation}. The inevitable loss of atoms in the ultracold regime, however, complicates the investigation of long-time dynamics and thereby hinders the analysis of long-range time crystalline order.
	
\begin{figure*}
	\centering
	\includegraphics[width=\linewidth]{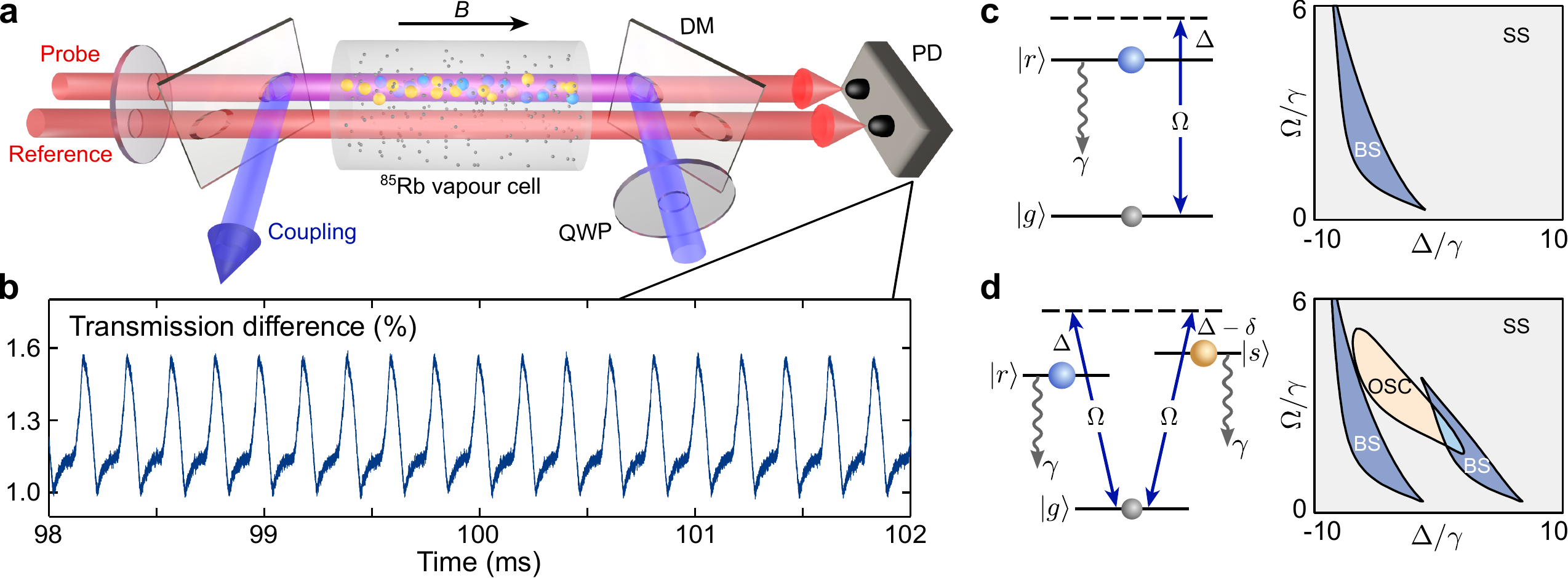}% Here is how to import EPS art
	\caption{ {\bf Experimental protocol and mean-field phase diagram.} {\bf a} Schematic of the experimental setup, where a probe and a reference beam are propagating in parallel through a room-temperature $^{85}$Rb vapour cell. The probe beam overlaps with a counterpropagating coupling beam, which is redirected by a dichroic mirror (DM) and can drive the atoms to Rydberg states. The quarter-wave plates (QWP) are used to control the polarization of the beams, and the transmission difference between the probe and the reference is detected by a balanced photon detector (PD). {\bf b} Single-run transmission difference for Rydberg states with a principal quantum number $n=75$. {\bf c} The left and the right panel respectively show the level scheme and the mean-field phase diagram for the case of a single Rydberg state $|r\rangle$ with $\chi=-16\gamma$, which supports a stationary phase (SS) and a bistable phase (BS). {\bf d} The left and the right panel respectively show the level scheme and the phase diagram for a system involving two Rydberg states $|r\rangle$ and $|s\rangle$ with $\chi=-16\gamma$ and $\delta=8\gamma$. Here, a limit cycle oscillating phase (OSC) is identified. The overlapping region between an OSC phase and a BS phase contains one stationary state and one stable limit cycle.}
	\label{fig:Fig1}
\end{figure*}

	Ensembles of Rydberg atoms represent a suitable platform for exploring many-body phenomena away from equilibrium emerging from coherent driving, dissipation, and long-range dipole-dipole interactions \cite{carr2013nonequilibrium,malossi2014full,ding2020phase,wu2021concise,horowicz2021critical,franz2022observation}. Such a Rydberg gas is well controllable and can be confined in a room-temperature vapour cell at virtually no atom loss. Here, we exploit this feature and report the experimental observation of long-range time crystalline order in a continuously driven Rydberg gas [see Fig.~\ref{fig:Fig1}{\bf a}]. In our experiment, the CTC manifests itself in limit cycle oscillations of the Rydberg atom density and the atomic dipole moment, which we probe directly by the transmission of light through the gas [see Fig.~\ref{fig:Fig1}{\bf b}]. We demonstrate that the time crystal originates from the competition between atomic excitations in different Rydberg states, and characterize the parameter regimes that support this phase. We verify the observation of a CTC by experimentally demonstrating the establishment of a long-range autocorrelation function and its robustness against noisy temporal perturbations.

	To understand the origin of the limit cycle in our experiment, we first consider the microscopic description of the driven-dissipative Rydberg gas. As illustrated in Fig.~\ref{fig:Fig1}{\bf c}, the applied laser fields generated a coherent coupling between the atomic ground state $|g\rangle$ and a Rydberg state $|r\rangle$, with a corresponding Rabi frequency $\Omega$ and frequency detuning $\Delta$. The loss of coherence is described by the decay of Rydberg population with a rate $\gamma$. The strong interaction between Rydberg states is governed by a Hamiltonian $\hat{H}_\mathrm{I}=\sum_{i\neq j}(V_{ij}/2)\hat{n}_i\hat{n}_j$, with $\hat{n}_i=|r_i\rangle\langle r_i|$ the local Rydberg density and $V_{ij}=C_6/|\mathbf{r}_i-\mathbf{r}_j|^6$ the van der Waals interaction between Rydberg atoms located at $\mathbf{r}_i$ and $\mathbf{r}_j$. In a thermal Rydberg gas, however, the atomic motion averages out the associated spatial correlations between Rydberg atoms and permits a mean-field treatment of the interaction \cite{carr2013nonequilibrium}, whereby the laser detuning $\Delta\rightarrow \Delta-\chi n_r$ acquires a dependence on the uniform Rydberg density $n_r=\langle\hat{n}_i\rangle$, with an effective nonlinearity strength $\chi$. The finite decay rate $\gamma$ usually relaxes the system to a stationary mixed state (SS), corresponding to the fixed point of the nonlinear optical Bloch equation. However, it has been shown that the interaction induced nonlinearity can cause a saddle-node bifurcation \cite{carr2013nonequilibrium}, through which a bistable stationary state emerges in a finite region of the parameter space. For an attractive interaction ($\chi<0$), such a bistable phase (BS) can occur at the negative detuning regime [see the right panel of Fig.~\ref{fig:Fig1}{\bf c}]. At its boundaries one finds a discontinuous nonequilibrium transition between distinct stationary states with low and high Rydberg-state concentration, respectively.
	
	The situation changes dramatically when more than one Rydberg states come into play. In order to illustrate this effect, we consider a minimal extension, in which one additional Rydberg state $|s\rangle$ is coupled with an identical Rabi frequency $\Omega$ but different detuning $\Delta-\delta$, [see  Fig.~\ref{fig:Fig1}{\bf d}]. These distinct Rydberg states can establish their respective bistable phases at different detunings $\Delta$ for a sufficiently large energy separation $\delta$. More importantly, their strong interactions generate nonlinear energy shifts, $E_\mathrm{NL}=\chi (n_r+n_s)$,
	that couple the dynamics of both Rydberg states (see Methods). The resulting competition can drive a Hopf bifurcation for $\chi\sim\delta$, whereby a non-stationary dynamical phase appears in between the two bistable regions, as illustrated in the right panel of Fig.~\ref{fig:Fig1}{\bf d}. In this non-stationary regime, the interaction can facilitate the excitation of one Rydberg state at the cost of the other, leading to limit cycle dynamics with persistent oscillations of the Rydberg densities without damping.
	
\begin{figure*}
	\centering
	\includegraphics[width=\linewidth]{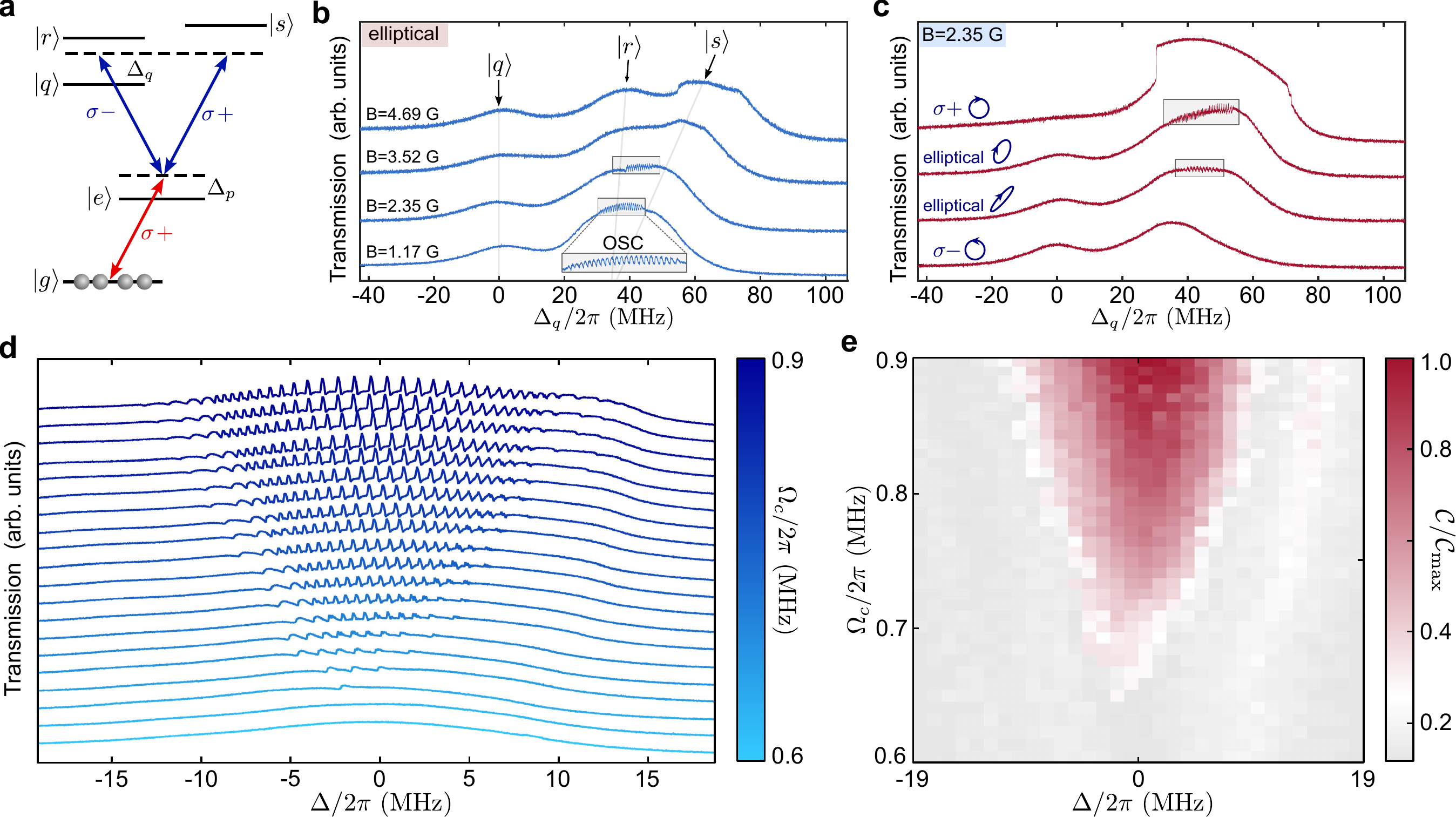}% Here is how to import EPS art
	\caption{{\bf Transmission spectrum and experimental phase diagram.} {\bf a} Energy level diagram for the Rydberg excitation scheme. The ground state $ |g\rangle=|5S_{1/2},F=3,m_F=3\rangle$ is coupled to an intermediate state  $ |e\rangle = |5P_{3/2},F=4,m_F=4\rangle$ by a $ \sigma^+ $ polarized 780 nm probe field, and three  Rydberg Zeeman sublevels  $|r\rangle = |nD_{5/2},m_J=1/2\rangle$, $|s\rangle = |nD_{5/2},m_J=5/2\rangle$, $|q\rangle = |nD_{3/2},m_J=1/2\rangle$ are further connected by a 480 nm coupling field. {\bf b} Scanned transmission spectrum for the different indicated magnetic fields at elliptical polarized coupling field. The Rabi frequencies for the probe and the coupling fields are $\Omega_p/2\pi=20$ MHz and $\Omega_c/2\pi=0.7$ MHz (for the transition to $|s\rangle$), and the intermediate-state detuning is $\Delta_p/2\pi=30$ MHz. {\bf c} Scanned transmission spectrum for the different indicated polarizations at a low magnetic field $ B=2.35 $ G. The Rabi frequencies are $\Omega_p/2\pi = 19$~MHz and $\Omega_c/2\pi = 1.3$~MHz (for a $\sigma^+$ circularly polarized coupling beam) with an intermediate-state detuning $\Delta_p/2\pi=94$~MHz. The scanning rate is $2\pi\times 2.84$~MHz/ms for {\bf b} and {\bf c}. {\bf d} Evolution of the scanned transmission spectrum at increasing Rabi frequency $\Omega_c$, performed with $n=75$, $\Omega_p/2\pi = 20~ \mathrm{MHz}$ and $\Delta_p/2\pi = 30~ \mathrm{MHz}$. The scanning rate is $ 2\pi\times 3.17 $ MHz/ms and the coupling field is linearly polarized. {\bf e} Extracted phase diagram with the oscillation contrast ratio $\mathcal{C}$ as the order parameter.}
	\label{fig:Fig2}
\end{figure*}	
	
	The experimental setup for observing such an oscillatory dynamics is depicted in Fig.~\ref{fig:Fig1}{\bf a}, where $^{85} $Rb atoms are trapped in a room-temperature vapour cell. The atoms are continuously excited to the Rydberg states via a two-photon process, in which a probe beam and a coupling beam counterpropagates with each other. Here, the probe and the coupling fields drive the ground state manifold $|5S_{1/2},F=3\rangle$ to the Rydberg manifold $|nD_J\rangle$ ($J=5/2,\ 3/2$) via intermediate states $|5P_{3/2},F=4\rangle$. The exact Zeeman level of the states can be specified by the polarization of the beams [see Fig.~\ref{fig:Fig2}{\bf a}]. In the limit cycle phase, the population of Rydberg states and the transition dipole moments are coupled with each other to form the persistent oscillation. The oscillation of the transition dipole moment between the ground and the excited state then results in an oscillating polarization on the probe transition, which can be directly monitored by the light transmission. To obtain a clear transmission signal, we use a calcite beam displacer to generate a reference beam parallel to the probe beam for differential measurement. The transmission signal for a high lying Rydberg state with a principal quantum number $n=75$ is shown in Fig.~\ref{fig:Fig1}{\bf b}, which exhibits a stable periodic oscillation pattern. In this single experiment, three Rydberg states close in energy are involved in the driving scheme \cite{su2022optimizing}, and should all participate in the synchronized oscillating dynamics.

	%\onecolumngrid
	%\begin{center}
	%\begin{figure}[h]
	%	\centering
	%	\includegraphics[width=\linewidth]{Fig2.pdf}% Here is how to import EPS art
	%	\caption{EIT signals at incresing $ \Omega_c $ from down to up. Zoom in to the oscillating region allows to see the rich forms of oscillations (in the grey shadow regimes). }
	%	\label{fig:Fig2}
	%\end{figure}
	%\end{center}
	%\twocolumngrid

	To confirm that the oscillation is indeed induced by the involvement of multiple Rydberg states, we reduce the principal quantum number to $n=69$ and apply a magnetic field ${B}$ to adjust the energy differences between distinct Zeeman levels. Then, with a careful choice of the polarization, we can study the dynamics mainly involving two Rydberg states close in energy, i.e., states $|r\rangle$ and $|s\rangle$ in Fig.~\ref{fig:Fig2}{\bf a}. The branching ratio of the coupling to these states can be tuned by the polarization of the coupling field \cite{su2022optimizing}. In order to identify the region displaying limit cycles, we perform a scanning spectroscopy measurement, in which the frequency of the coupling beam is slowly scanned near the two-photon resonance with a fixed intermediate-state detuning $\Delta_p$. In the scanning process, the magnetic insensitive Rydberg level $|q\rangle = |69D_{3/2},m_J=1/2\rangle$ separated from the two target states can also be laser coupled, and is chosen as a reference state with a detuning $\Delta_q$.
	
	The observed transmission spectrum for different magnetic fields and polarizations are shown in Figs.~\ref{fig:Fig2}{\bf b} and \ref{fig:Fig2}{\bf c}. First, we set the polarization of coupling beam to be elliptical, by which both Rydberg states can be excited, and they are distinguishable at a large magnetic field $B=4.69\ \mathrm{G}$ [see Fig.~\ref{fig:Fig2}{\bf b}]. Crucially, oscillations appear only in the case of a relatively small magnetic field, where $ |r\rangle $ and $ |s\rangle $ are close enough to induce the competition. Next, we fix the magnetic field at a small value $B = 2.35~\mathrm{G}$ and vary the polarization of the coupling light [see Fig.~\ref{fig:Fig2}{\bf c}]. We find the oscillating phase only exists at an elliptical polarization, but disappears for a perfectly circularly polarized case $\sigma^-$ or $\sigma^+$, where only one of the Rydberg states $|r\rangle$ or $|s\rangle$ is excited. Based on this experimental evidence, we can conclude that it is the competition between multiple Rydberg levels facilitating the limit cycle phase, in agreement with the mean-field prediction.

		\begin{figure*}
	\centering
	\includegraphics[width=\linewidth]{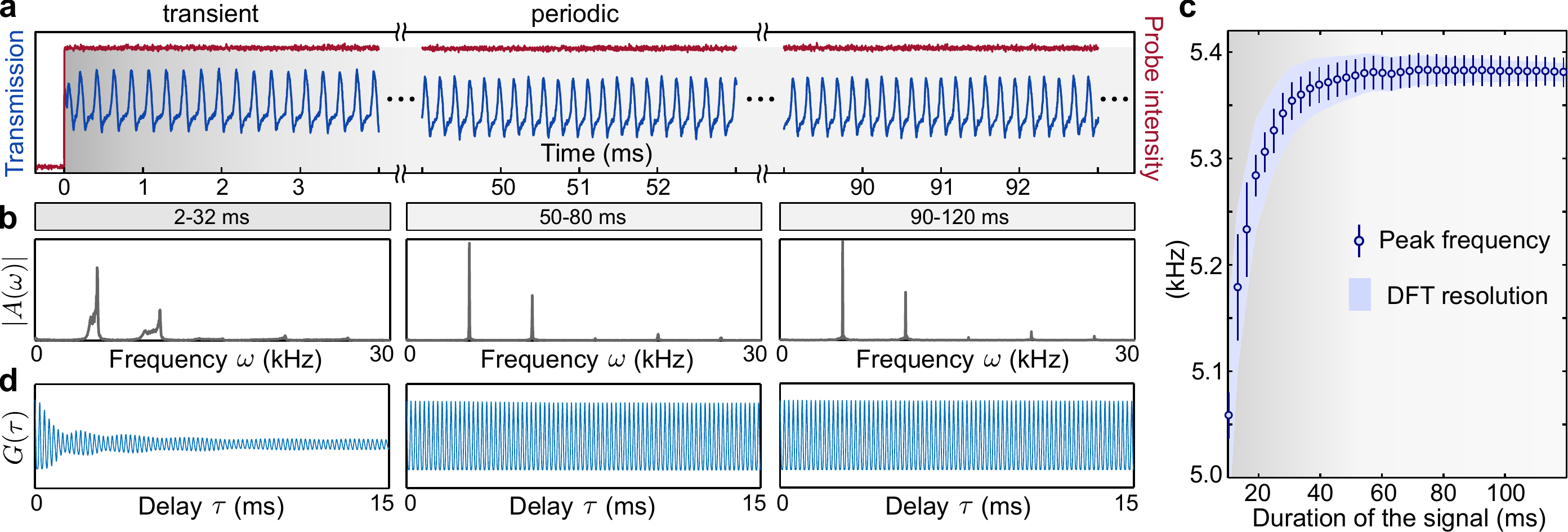}% Here is how to import EPS art
	\caption{{\bf Establishment of the long-range temporal order.}  {\bf a} shows a single-shot realization of the quench dynamics, where the blue line represents the sustained oscillating signal of the transmission, and the red lines represent the intensity of the probe field monitored by an independent PD. {\bf b} displays the discrete Fourier-transform of the time-dependent transmitted power recorded in {\bf a}, performed for different time windows 2-32~ms, 50-80~ms, and 90-120~ms (from left to right). {\bf c} Evolution of the peak frequency by varying duration of the signal starting from 2~ms. The error bar corresponds to the standard deviation of 100 independent realizations and the light blue shading represents the frequency resolution of the DFT. {\bf d} shows the ACF $G(\tau)$ for the time windows considered in {\bf b}. The coupling scheme is the same as Fig.~\ref{fig:Fig2}{\bf d}, with $\{\Omega_p, \Omega_c\}/2\pi = \{18,0.9\}$ MHz.}
	\label{fig:Fig3}
\end{figure*}	
	
	In addition to multiple Rydberg components, sufficiently large Rabi frequencies of the probe ($\Omega_{p}$) and the coupling ($\Omega_{c}$) are also necessary to induce the limit cycle. Figure \ref{fig:Fig2}{\bf d} displays the transmission spectrum at different $\Omega_c$ with a large and fixed $\Omega_p$, from which rich oscillating patterns can be identified. First, we note that the oscillation completely disappears if the Rabi frequency is too small. Then, weak and long-period oscillations come into appearance when $\Omega_c$ slightly exceeds a certain critical point. As the Rabi frequency is further increased, the amplitude of the oscillation as well as its range increases. Choosing the oscillating contrast ratio $\mathcal{C}$ as the order parameter, we can extract an approximate phase diagram from the transmission spectrum, as shown in Fig.~\ref{fig:Fig2}{\bf e}, where limit cycle phase is revealed by a nonvanishing contrast. The measured phase diagram indicates that the limit cycle phase exists over a wide parameter regimes and is not sensitive to experimental conditions, such as laser power and frequency, etc.
	
	Having identified parameter regions that support oscillatory states, we can now study their properties in relation to the physics of time crystals. Specifically, we investigate a quench dynamics, where the probe field is suddenly turned on and then held at a constant strength for a very long time. As a typical realization shown in Fig.~\ref{fig:Fig3}{\bf a}, the elementary oscillation pattern is rapidly established within $\sim 200~ \mu\mathrm{s}$, and then sustained in the entire driving window of $125~\mathrm{ms}$. A detailed analysis of the signal reveals that the early oscillations have a slowly drifting frequency during the initial transient dynamics, which evolve to periodic oscillations with a steady frequency. Figure \ref{fig:Fig3}{\bf b} displays the discrete Fourier transform (DFT) of the transmission signal at different stages. For the time window 2-32~ms, the Fourier spectrum has several broad peaks, indicating the absence of a well-defined periodicity. In contrast,
	the spectrum features equidistant narrow Fourier peaks for the time window 50-80~ms and keeps its shape
	at later times 90-120~ms, suggesting that a stable periodic pattern has been reached. We then perform 100 independent measurements and study evolution of the oscillation frequencies. As shown in Fig.~\ref{fig:Fig3}{\bf c}, the fundamental frequency drifts towards higher values in the transient region, and then converges to a stable one. These repeated realizations of the quench dynamics confirm that the above evolution towards a stable oscillating periodicity is not a consequence of fluctuations of the experimental conditions but rather an inherent process of the interacting atoms.
	
	We further extract the autocorrelation function (ACF) $G(\tau)=\int dt I(t)I(t+\tau)$, where $I(t)$ is the shifted transmission signal with zero mean [$\int dtI(t)=0$]. The ACF is the classical estimation of the two-time quantum correlation function \cite{medenjak2020isolated,guo2022quantum,greilich2023continuous}, which can quantify temporal correlations of the dynamics. Figure \ref{fig:Fig3}{\bf d} shows the ACF at the same stages as in Fig.~\ref{fig:Fig3}{\bf b}. In the transient stage, the envelope of the ACF exhibits approximately an algebraic decay with delay $\tau$, in accordance with a quasi long-range order. Remarkably, as the system enters the stable periodic stage, the ACF shows a persistent oscillation with an almost constant amplitude within the considered range of the delay $\tau$ ($\sim 80$ periods). The nondecaying behavior of the ACF here corroborates the establishment of a true long-range order (LRO), which is a strong evidence of the time crystal. 

	\begin{figure*}
	\centering
	\includegraphics[width=\linewidth]{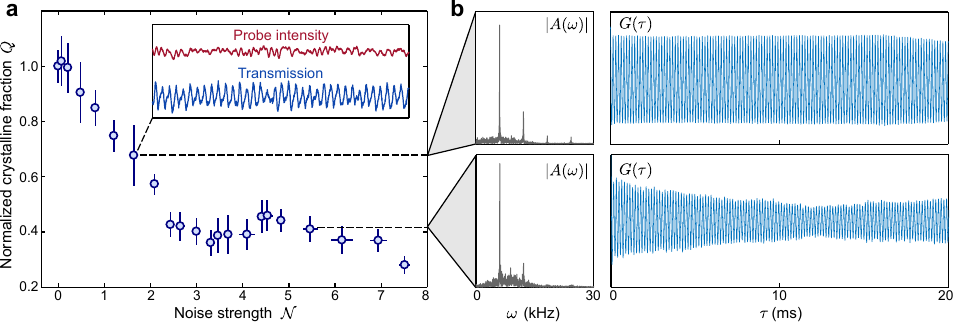}% Here is how to import EPS art
	\caption{{\bf Robustness against temporal perturbations and melting of the observed time crystal.} {\bf a} Normalized crystalline fraction as a function of the noise strength $\mathcal{N}$, defined as $ \mathcal{N}=\sum_{\omega}|P_{n}(\omega)|/\sum_{\omega}|P_{0}(\omega)|-1 $, where $P_{n}(\omega) $ and $ P_{0}(\omega) $ correspond to the single-sided amplitude spectrum of the monitored probe intensity with and without noise, respectively. The vertical and horizontal error bars respectively represent the standard deviation of the normalized crystalline fraction $Q$ and the noise strength $\mathcal{N}$ of 10 independent realizations. The inset illustrates the transmission signal (blue) and the monitored probe intensity (red) at the data point indicated by the dashed line. {\bf b} The left and the right panels respectively correspond to the modulus of the amplitude spectrum $|A(\omega)|$ and the autocorrelation function $G(\tau)$ at the data points indicated by the dashed line. The experimental parameters are the same as in Fig.~\ref{fig:Fig3}.}
	\label{fig:Fig4}
\end{figure*}
	
	As a defining property, a CTC is predicted to be rigid, i.e., robust against temporal perturbations. To test the rigidity of the observed limit cycle, we add intensity noise on the probe light during the quench dynamics. For a small but finite noise strength $\mathcal{N}$, we note that the ordered oscillation of the transmission signal will not suddenly disappear, but preserves the basic pattern found in the noise-free case [see the inset of Fig.~\ref{fig:Fig4}{\bf a}]. With increasing noise strength, the random noise gradually dominates over the ordered oscillation. To quantify such a melting process of the CTC, we introduce a relative crystalline fraction \cite{kongkhambut2022observation}, defined as $Q = {\sum_{|\omega-\omega_c|\leq\delta\omega} |A(\omega)}|/{\sum_{\omega} |A(\omega)}|$, where $\omega_c$ and $\delta \omega$ respectively denote the fundamental frequency and the resolution of the DFT amplitude $A(\omega)$. As shown in Fig.~\ref{fig:Fig4}{\bf a}, the measured crystalline fraction drops almost linearly with the noise strength $\mathcal{N}$ in the initial stage and then holds within a certain range of strong noise strengths. To investigate the temporal correlation of the melting time crystal, we extract the DFT spectrum and the ACF for measurements performed in different stages [see Fig.~\ref{fig:Fig4}{\bf b}]. In the linear drop zone, the DFT spectrum still has equally separated peaks embedded in a noisy background. The oscillation of the ACF remains persistent before a slight drop at a very large delay $\tau$, signaling a preserved LRO. Entering the strong noise region, while the DFT amplitude spectrum has a sharp peak at the fundamental frequency, high-frequency components are erased by the noise. Meanwhile, the noise destroys the LRO, as revealed by the decay of the autocorrelation function for $\tau>0$. These observations demonstrate that the long-range temporal correlation can be partially maintained during the melting, confirming that the observed dissipative time crystal is robust to temporal fluctuations of the continuous driving.
	
	In summary, we experimentally demonstrate dissipative time crystalline order in a room-temperature Rydberg atom ensemble, which is directly observed by the oscillatory transmission of the probe beam. The observed oscillation originates from the simultaneous coupling to distinct Rydberg states, by which the competition between different Rydberg components facilitates the emergence of limit cycles. Importantly, the persistent oscillation has true long-range temporal order and is robust against noisy perturbations, fulfilling the fundamental criteria of a CTC. By extending the accessible range of the physical parameters, e.g., the Rabi frequency and the atomic density, it will be interesting to measure the complete phase diagram including bistability phases, which can be distinguished via hysteresis measurements \cite{carr2013nonequilibrium}. By employing a periodic driving, the current setup also holds promise for the study of a discrete time crystal \cite{else2016floquet,gong2018discrete,lazarides2020time,cabot2022metastable,pal2018temporal,cosme2019time,gambetta2019discrete,tuquero2022dissipative}. Realization of the model in a Rydberg-atom tweezer array could allow for further studies of the time-crystalline order, such as the behavior in different spatial dimensions, the role of quantum fluctuations and correlations, as well as their scaling with the system size. Our work provides a realistic interacting many-body system for systematically investigating the dissipative time crystal, and opens up a new route to quantum synchronization \cite{khasseh2019many,buvca2022algebraic} and sensing \cite{Ilias2022criticality,cabot2023continuous}.
	
	{\it Note added}. After completion of this work, we became aware of two recent reports \cite{ding2023ergodicity,wadenpfuhl2023emergence}. Ref.~\cite{ding2023ergodicity} investigates transient oscillations due to inhomogeneous Rydberg excitation. Ref.~\cite{wadenpfuhl2023emergence} studies the synchronization of atoms with different velocities and observes oscillations in the transmission spectrum as a function of the control-field detuning in a heated vapor cell away from EIT resonance.
	
	\begin{acknowledgments}
		We acknowledge valuable discussions with Drs. Klaus M{\o}lmer, Sebastian Hofferberth, Yanhong Xiao, Yaofeng Chen, Feng Chen, Yuanjiang Tang, Hadi Yarloo, and Huachen Zhang, as well as the support by Prof. Luyi Yang. This work is supported by the National Key R$\&$D Program of China (Grant No.~2018YFA0306504 and No.~2018YFA0306503), the National Natural Science Foundation of China (NSFC) (Grant No. 92265205 and 12361131576), and the Innovation Program for Quantum Science and Technology (2021ZD0302100). F. Yang and T. Pohl acknowledge the support from Carlsberg Foundation through the ``Semper Ardens'' Research Project QCooL and from the Danish National Research Foundation (DNRF) through the Center of Excellence ``CCQ'' (Grant No.~DNRF156). T. Pohl acknowledges the support from the Austrian Science Fund (FWF) [10.55776/COE1] and the European Union - NextGenerationEU, and the Horizon Europe ERC synergy grant SuperWave (grant no. 101071882).
	\end{acknowledgments}
	
	\noindent	
	\textbf{Author contributions}\par
	\noindent
	X.W. constructed the initial experimental set up and observed the primary phenomenon, while Z.W. and X.L. gave crucial assistance.  Z.W., R.G., and X.W. performed the experiments and analysed the data after thorough discussions with F.Y. The theoretical model was proposed by X.W., Z.W., F.Y., and T.P. The theoretical analysis and numerical simulations are conducted by F.Y. All authors discussed the results and contributed to the manuscript. X.L., T.P., and L.Y. supervised the study.

	\noindent \textbf{Competing interests}\par\noindent The authors declare no competing interests.
	
	%\bibliographystyle{mybst}
	%\bibliography{ref}
	
	%\newpage

	%\setcounter{equation}{0}
	%\setcounter{figure}{0}
	%\renewcommand{\figurename}{\textbf{Extended Data Figure}}

	\newpage
	
	\setcounter{equation}{0}
	\setcounter{figure}{0}
	\renewcommand{\figurename}{\textbf{Extended Data Figure}}

	\vspace{0pt}
	\section*{Methods}
	\noindent\textbf{Mean-field description of the dynamics}\par
\begin{figure*}[h]
		\centering
		\includegraphics[width=\linewidth]{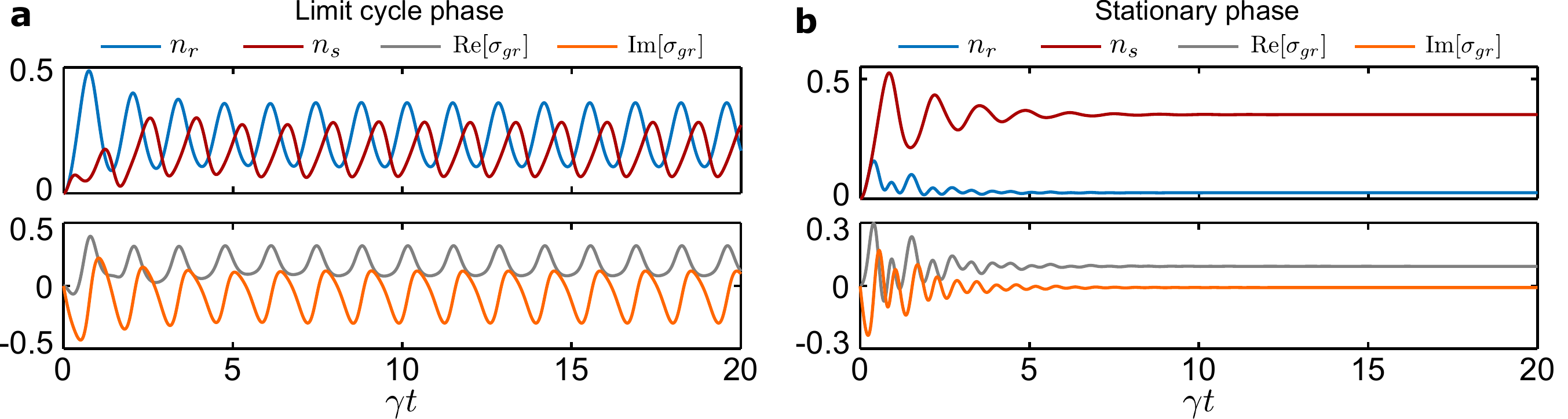}% Here is how to import EPS art
		\caption{ {\bf Mean-field dynamics. } The calculations are performed with $\Omega=3\gamma$, $\chi=-16\gamma$, and $\delta=8\gamma$. For the limit cycle phase shown in {\bf a} with $\Delta=-3\gamma$, the order parameters exhibit persistent oscillations. For the stationary phase shown in {\bf b} with $\Delta=3\gamma$, the order parameters converge to steady-state values.}
		\label{fig:FigA1}
	\end{figure*}

\noindent
	To qualitatively understand the observed oscillation, we can consider a simple V-type three-level model shown in Fig.~\ref{fig:Fig1}{\bf d}. Then, the time translation invariant Hamiltonian (in the rotating frame) of the model is given by
	\begin{align}
		\hat{H}=\ &\frac{\Omega}{2}\sum_i\left(\hat{\sigma}_i^{gr}+\hat{\sigma}_i^{gs}+\mathrm{H.c.}\right)-\sum_i\left(\Delta_r\hat{n}_i^r+\Delta_s\hat{n}_i^s\right)\nonumber\\
		&+\frac{1}{2}\sum_{i\neq j}V_{ij}\left(\hat{n}_i^r\hat{n}_j^r+2\hat{n}_i^r\hat{n}_j^s+\hat{n}_i^s\hat{n}_j^s\right),
	\end{align}
	where $\hat{\sigma}_i^{\alpha\beta}=|\alpha_i\rangle\langle\beta_i|$ is the atomic transition operator ($\alpha,\beta=g,r,s$), and $\hat{n}_i^\alpha=|\alpha_i\rangle\langle \alpha_i|$ ($\alpha=r,s$) denotes the local Rydberg density. Here, we assume an equal interaction strength $V_{ij}$ for different Rydberg levels $|r\rangle$ and $|s\rangle$. Taking the Rydberg decay into consideration, the evolution of an operator $\hat{\mathcal{O}}$ is governed by
	\begin{equation}
		\partial_t\langle\hat{\mathcal{O}}\rangle=\langle i[\hat{H},\hat{\mathcal{O}}]+\mathcal{L}_r^*[\hat{\mathcal{O}}]+\mathcal{L}_s^*[\hat{\mathcal{O}}]\rangle,
	\end{equation}
	where $\mathcal{L}_\alpha^*[\hat{\mathcal{O}}]=(\gamma/2)\sum_i\left(2\hat{\sigma}_i^{\alpha g}\hat{\mathcal{O}}\hat{\sigma}_i^{g\alpha}-\{\hat{n}_i^{\alpha},\hat{\mathcal{O}}\}\right)$ is the Lindbladian describing the decay of the Rydberg state $|\alpha\rangle$ ($\alpha=r,s$). The equations of motion for the first moments are then determined by
	\begin{align}
		\partial_t\langle\hat{n}_{j}^{r}\rangle =&\ i\frac{\Omega}{2}[\langle\hat{\sigma}_{j}^{gr}\rangle-\langle\hat{\sigma}_{j}^{rg}\rangle]-\gamma\langle\hat{n}_{j}^{r}\rangle,  \nonumber \\
		\partial_t\langle\hat{n}_{j}^{s}\rangle =&\   i\frac{\Omega}{2}[\langle\hat{\sigma}_{j}^{gs}\rangle-\langle\hat{\sigma}_{j}^{sg}\rangle]-\gamma\langle\hat{n}_{j}^{s}\rangle,  \nonumber\\
		\partial_t\langle\hat{\sigma}_{j}^{gr}\rangle =&\  i\frac{\Omega}{2}[\langle\hat{n}_{j}^{r}\rangle-\langle\hat{\sigma}_{j}^{gg}\rangle+\langle\hat{\sigma}_{j}^{sr}\rangle]+i\left(\Delta_r+i\frac{\gamma}{2}\right)\langle\hat{\sigma}_{j}^{gr}\rangle   \nonumber\\
		&-i\sum_{k\neq j}V_{jk}[\langle\hat{n}_{k}^{r}\hat{\sigma}_{j}^{gr}\rangle+\langle\hat{n}_{k}^{s}\hat{\sigma}_{j}^{gr}\rangle
		], \nonumber\\	
		\partial_t\langle\hat{\sigma}_{j}^{gs}\rangle =&\ i\frac{\Omega}{2}[\langle\hat{n}_{j}^{s}\rangle-\langle\hat{\sigma}_{j}^{gg}\rangle+\langle\hat{\sigma}_{j}^{rs}\rangle]+i\left(\Delta_s+i\frac{\gamma}{2}\right)\langle\hat{\sigma}_{j}^{gs}\rangle   \nonumber\\
		&-i\sum_{k\neq j}V_{jk}[\langle\hat{n}_{k}^{s}\hat{\sigma}_{j}^{gs}\rangle+\langle\hat{n}_{k}^{r}\hat{\sigma}_{j}^{gs}\rangle
		],    \nonumber \\
		\partial_t\langle\hat{\sigma}_{j}^{rs}\rangle =&\ i\frac{\Omega}{2}[\langle\hat{\sigma}_{j}^{gs}\rangle-\langle\hat{\sigma}_{j}^{rg}\rangle]-i(\Delta_r-\Delta_s-i{\gamma})\langle\hat{\sigma}_{j}^{rs}\rangle,   \nonumber
	\end{align}
	In the mean-field treatment \cite{lee2011antiferromagnetic,qian2012phase,vsibalic2016driven,he2022superradiance}, we neglect the correlation between different atoms, which allows for a factorization of the high-order moments, e.g., $\langle\hat{n}_{k}^{r}\hat{\sigma}_{j}^{gr}\rangle=\langle\hat{n}_{k}^{r}\rangle\langle\hat{\sigma}_{j}^{gr}\rangle$. Assuming a uniform spatial distribution \cite{marcuzzi2014universal}, e.g., $\langle\hat{n}_{j}^{r}\rangle=n_r$, we arrive at the mean-field equations:
	\begin{align}
		\dot{n}_r &=i\frac{\Omega}{2}({\sigma}_{gr}-{\sigma}_{rg})-\gamma{n}_{r},  \nonumber \\
		\dot{n}_{s} &=  i\frac{\Omega}{2}({\sigma}_{gs}-{\sigma}_{sg})-\gamma{n}_{s}, \nonumber \\
		\dot{\sigma}_{gr} &= i\frac{\Omega}{2}(2{n}_{r}+{n}_{s}+{\sigma}_{sr}-1)+i\left(\Delta_r-E_\mathrm{NL}+i\frac{\gamma}{2}\right){\sigma}_{gr},   \nonumber \\	
		\dot{\sigma}_{gs} &=i\frac{\Omega}{2}(2{n}_{s}+{n}_{r}+{\sigma}_{rs}-1)+i\left(\Delta_s-E_\mathrm{NL}+i\frac{\gamma}{2}\right){\sigma}_{gs},   \nonumber    \\
		\dot{\sigma}_{rs} &=i\frac{\Omega}{2}({\sigma}_{gs}-{\sigma}_{rg})-i(\Delta_r-\Delta_s-i{\gamma}){\sigma}_{rs},   \nonumber
	\end{align}
	where $E_\mathrm{NL}=\chi({n}_{r}+{n}_{s})$ is the nonlinear energy shift with $\chi=\sum_{k\neq j}V_{jk}$. The stable behavior of the system is determined by the property of the fixed points of the above nonlinear Bloch equations. With a standard stability analysis \cite{strogatz2018nonlinear}, we can analyze types of bifurcations and obtain the phase diagram shown in Fig.~\ref{fig:Fig1}{\bf d}, where we set $\Delta_r=\Delta$ and $\Delta_s=\Delta-\delta$. In Extended Data Fig.~\ref{fig:FigA1}, we calculate the mean-field dynamics in different regimes. When the system is in the limit cycle phase, the mean-field components exhibit persistent periodic oscillations after an initial synchronization [see Extended Data Fig.~\ref{fig:FigA1}{\bf a}]. Here, oscillation of the imaginary part of the transition dipole $\sigma_{gr}$ causes an oscillating absorption coefficient of the probe field, which can be directly monitored by the transmission signal as in our experiment. Instead, if the system is in the stationary phase [see Extended Data Fig.~\ref{fig:FigA1}{\bf b}], all observables will eventually converge to their steady-state values.
	
\begin{figure*}[h]
		\centering
		\includegraphics[width=0.6\linewidth]{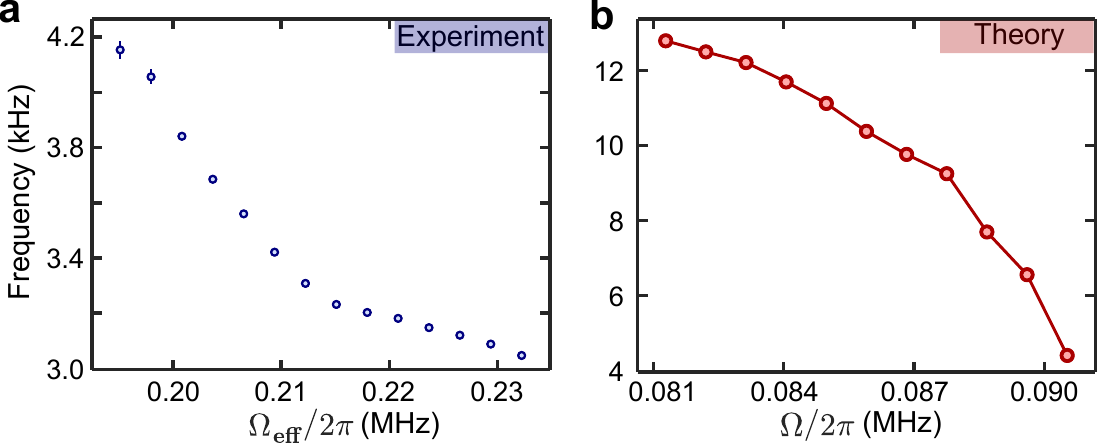}% Here is how to import EPS art
		\caption{ {\bf Comparison between experiment and theory.  }{\bf a} Experimentally extracted oscillation frequency as a function of the effective two-photon Rabi frequency $\Omega_\mathrm{eff}=\Omega_c\Omega_p/2\Delta_p$. In the experiment, we only vary the intensity of the coupling field ($\Omega_c$), and keep all other conditions the same as in Fig.~\ref{fig:Fig2}{\bf d}. {\bf b} Theoretically predicted oscillation frequency as a function of the Rabi frequency $\Omega$. The parameters chosen are close to the experiment. For example, the decay rate is evaluated to be $\gamma/2\pi=18.0~$kHz, consisting of the contribution from the Rydberg spontaneous decay ($\gamma_d$) and the transient time broadening $\gamma_t=\bar v_\perp/D$ with $D$ the diameter of the beam and $\bar v_\perp=\sqrt{{\pi k T}/{2m}}$ the transverse mean velocity. The other parameters $\delta/2\pi=0.306~$MHz, $\Delta/2\pi=-0.186~$MHz, and $\chi/2\pi=-0.951~$MHz are also typical in our experiment.}
		\label{fig:FigA2}
	\end{figure*}

	The primary aim of the mean-field model is to develop an intuitive understanding of the experiment and to show that the inclusion of multiple Rydberg-levels promotes the observed persistent oscillations. Here, we also compare the model prediction with the experiment on a more quantitative level. We focus specifically on the emerging frequency of the oscillation and its dependence on the system parameters. For typical parameters of the experiment, we identify oscillatory solutions with high as well as low frequencies. While high oscillatory usually overestimate the frequency by more than one order of magnitude, the slow oscillatory solutions can yield typical frequencies on the same order as in our experiment and exhibit the same dependence on the system parameters. As illustrated in Extended Data Fig.~\ref{fig:FigA2}{\bf a}, the measured oscillation frequency decreases significantly with a slight increase of the driving Rabi frequency, in agreement with the mean-field prediction shown in Extended Data Fig.~\ref{fig:FigA2}{\bf b}. Such a behavior is not immediately obvious, as one would normally expect a positive correlation between the oscillation frequency and the driving Rabi frequency, e.g., in a single-atom case.
	
	We therefore believe that our model captures the essential physics of the observed time crystal formation and can serve as a starting point for developing a more complicated theory, where more details (e.g., incorporation of the intermediate state, the random thermal motion, Doppler effects induced by atoms with different velocities, and the laser-beam inhomogeneity) can be taken into account to yield more quantitative predictions. \\

	\noindent\textbf{Experimental details}\par
\begin{figure*}[h]
		\centering
		\includegraphics[width=0.96\linewidth]{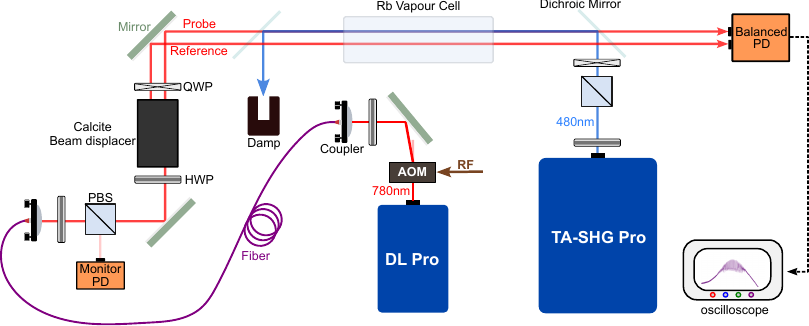}% Here is how to import EPS art
		\caption{ {\bf A detailed schematic diagram of the experimental setup. } The 780 nm probe and reference beams generated from a calcite beam displacer are propagating in parallel through a room-temperature Rb vapor cell, with the probe beam overlapping with a counterpropagating coupling beam. Their transmission signals are detected on a balanced photon detector (PD) for differential measurement. The output intensity of the 780 nm laser from the fiber (monitored by an independent PD) is controlled by an AOM, whose RF driving is modulated by a waveform generator. QWPs are used to control the polarization of the lasers illuminating the atoms.}
		\label{fig:FigA3}
	\end{figure*}

\noindent
	The detailed experimental setup is depicted in Extended Data Fig.~\ref{fig:FigA3}. Our experiments are performed in a 7.5-cm-long room-temperature Rb vapour cell \cite{miller2016radio,ripka2018room,li2022telecom} with natural abundance, where the proportions of $ ^{85} $Rb and $ ^{87} $Rb isotopes are about 72.2\% and 27.8\%. The laser sources of 780 nm and 480 nm light fields are [Toptica DL pro @780 nm] and [Toptica TA-SHG pro @480 nm], and both of them can work in the frequency-scanning mode or the frequency-locking mode. For 780 nm laser, we use an acoustic optical modulator (AOM) before the coupler of the fiber to control the output laser intensity, which is monitored by an independent PD. Then, a calcite beam displacer (Thorlabs BD40) is used to generate two parallell 780 nm beams for differential measurements, one of which is chosen as the probe beam (overlapped with a 480 nm coupling beam) with the other being a reference. The maximum powers of the probe ($ 1/e^2 $ waist radius $ 1000\ \mu m $) and the coupling ($ 1/e^2 $ waist radius $ 900\ \mu m $) beams are approximately $ 1.2\ \mathrm{mW} $ and $ 460\ \mathrm{mW} $, with which the maximum Rabi frequencies are approximately $ \Omega_p^{(\mathrm{max})}\sim 2\pi\times  20.5\ \mathrm{MHz}$ and $ \Omega_c^{(\mathrm{max})}\sim 2\pi\times  1.9\ \mathrm{MHz}$ (dependent on the laser polarization and the specific Zeeman level). The transmission spectrum is obtained by scanning the frequency of the coupling field (the detuning $\Delta_c(t)$) at a fixed intermediate-state detuning $ \Delta_p $ around the two-photon resonance $ \Delta_p+\Delta_c=0 $.

	For the measurement of the quench dynamics, the frequencies of both laser sources are locked to an ultralow expansion (ULE) cavity with the help of the Pound-Drever-Hall (PDH) technique. Then, by controlling the radio frequency (RF) driving on the AOM, we can turn on and off the 780 nm output laser abruptly within $ 2\ \mu$s, during which the 480 nm laser is always kept on. When exploring the rigidity of the dissipative time crystal, the intensity noise added to the probe field is achieved by using a waveform generator (Keysight 33600A Series) to modulate the RF amplitude.
	
	To adjust the energy spacing of different Rydberg Zeeman sublevels, we use a homemade 7.5-cm-long coil with 35 windings to generate a homogeneous external magnetic field parallel to the laser beams. Meanwhile, quarter-wave plates (QWP) are used to control the polarizations of the 780 nm and 480 nm light fields, with which we can tune the coupling strengths to different Zeeman levels (see Fig.~\ref{fig:Fig2} of the main text). 
	
	While the data shown in this paper is based on the excitation of $^{85}$Rb isotopes, similar oscillatory behavior is also observed for $^{87}$Rb isotopes, indicating that the oscillating dynamics is a universal phenomenon in Rydberg atomic ensembles. However, due to the lower natural abundance of $^{87}$Rb isotopes, the limit cycle usually requires a larger Rabi frequency or an additional heating to increase the Rydberg excitation density.\\

	\noindent\textbf{Influence of the principal quantum number}\par
\begin{figure*}[h]
		\centering
		\includegraphics[width=0.5\linewidth]{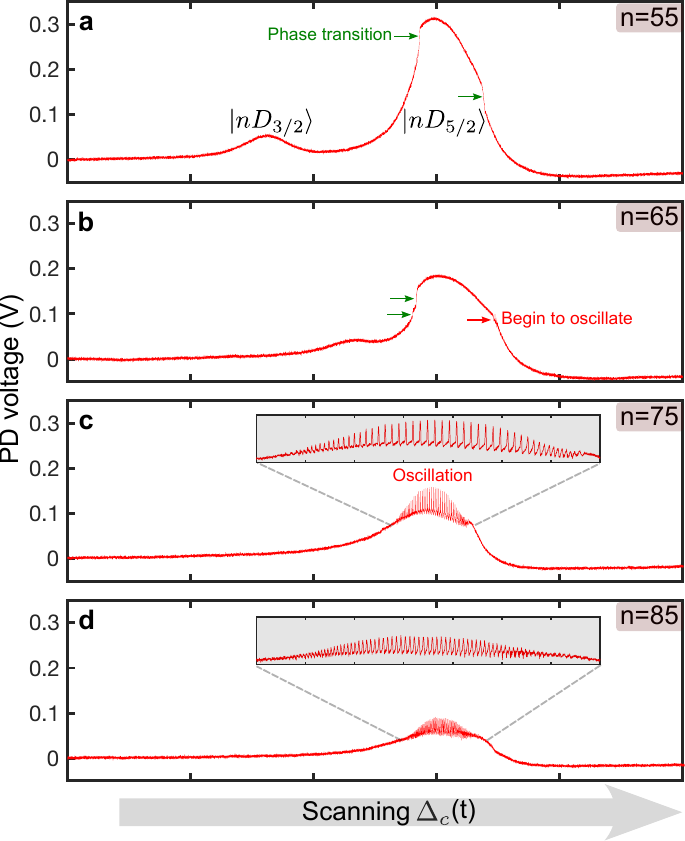}% Here is how to import EPS art
		\caption{{\bf Transmission spectrum at different principal quantum number $n$.} As $n$ increases, $ |nD_{3/2}\rangle $ and $ |nD_{5/2}\rangle $ states move closer, and their corresponding transmission peaks begin to merge. Oscillation appears and is enhanced for a sufficiently large $n$.}
		\label{fig:FigA4}
	\end{figure*}

\noindent
	It is necessary to excite atoms to a high principal quantum number $ n $ for the observation of oscillatory signals.  Figures~\ref{fig:FigA4}{\bf a}-{\bf d} show the transmission spectrum at different $ n $, where the probe (coupling) beam is elliptically (linearly) polarized with no external magnetic field. At a relatively small $ n=55 $, as expected, there are only two distinct transmission peaks, representing the resonant positions or energy levels of two Rydberg $nD$-state manifolds with total angular momentum quantum numbers $ J=3/2 $ and $ J=5/2 $. The peak of the latter is higher due to the larger dipole matrix element between the Rydberg state and the intermediate state. In other words, the Rabi frequency $ \Omega_c $ of coupling beam for $ |55D_{5/2}\rangle $ is larger than $ |55D_{3/2}\rangle $. Consequently, the Rydberg population is also higher for $ |55D_{5/2}\rangle $, leading to non-equilibrium dynamical phase transitions in bistable regions where the signal abruptly changes (green arrows).
	
	When $n$ is increased to 65 [see Extended Data Fig.~\ref{fig:FigA4}{\bf b}], the overall peaks become lower due to the smaller coupling Rabi frequency which scales as $ \sim n^{-3/2} $. The energy difference between two Rydberg states also becomes smaller. Here, multiple phase transitions in the left side of $ |65D_{5/2} \rangle$ are identified, and the emergent oscillatory signal appears in the right side of $ |65D_{5/2} \rangle$ (red arrow). Dramatic changes to the above features occur when $ n $ is further increased [see Extended Data Fig.~\ref{fig:FigA4}{\bf a}-{\bf d}]. As $ |nD_{3/2}\rangle $ and  $ |nD_{5/2}\rangle $ become closer, the two Rydberg states come to merge and the oscillation signal gets amplified over a wide range of the detuning $\Delta_c$. In general, the appearance and the enhancement of the oscillating dynamics with an increasing $n$ can be attributed to: (i) the stronger Rydberg interactions, and (ii) the participation of more Rydberg states (within $|nD_{3/2}\rangle $ and $|nD_{5/2}\rangle $ manifold) in the dynamics.\\
	
	\noindent \textbf{Long-term stability and random phase fluctuation.}\par
\begin{figure*}[h]
		\centering
		\includegraphics[width=0.6\linewidth]{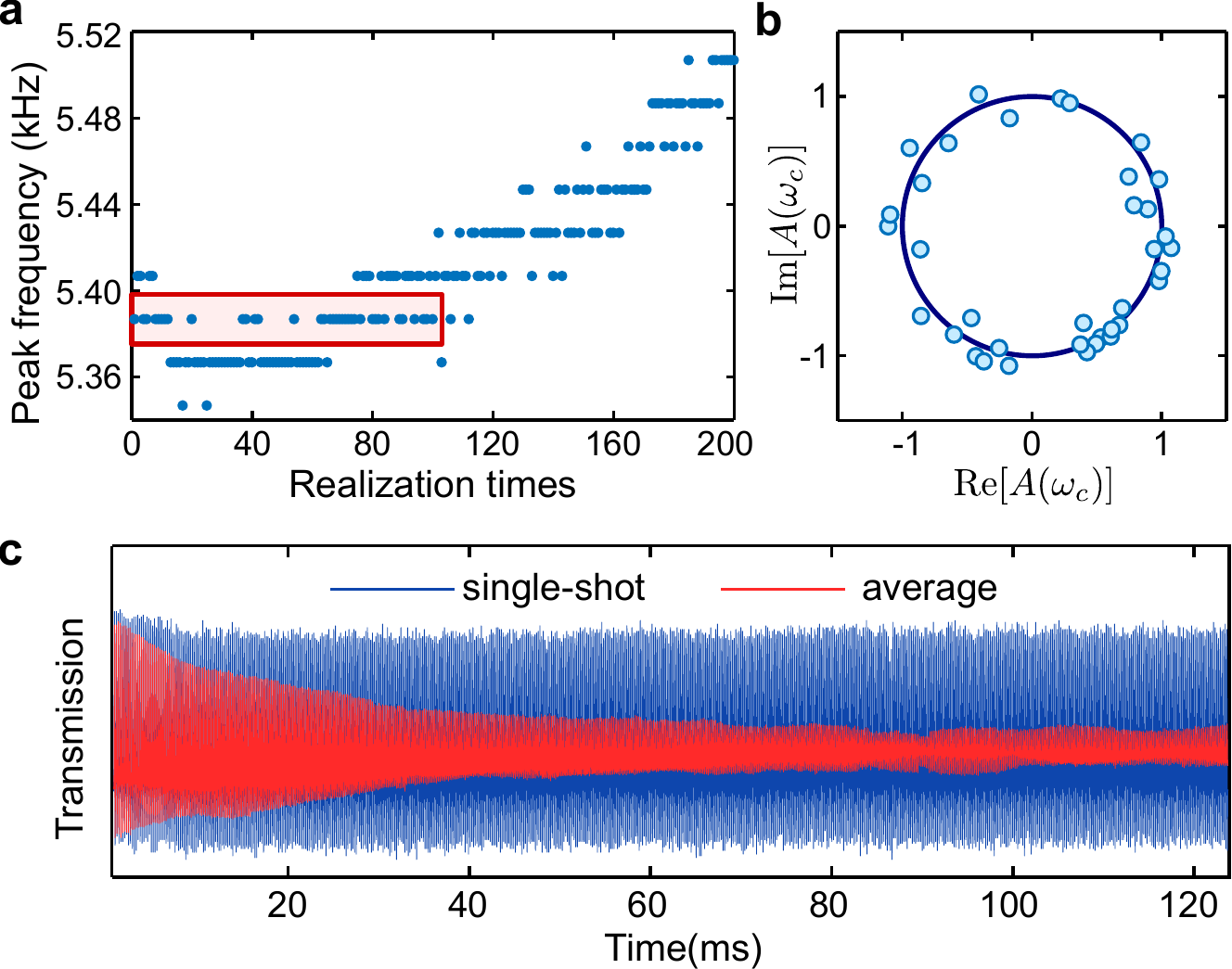}% Here is how to import EPS art
		\caption{ {\bf Long-term stability of the oscillation frequency and breaking of continuous time translation symmetry. }  {\bf a} The long-term frequency drift in 200 independent realizations of the quench dynamics by extracting peak frequencies in the stable periodic region (50-100~ms). The data encircled by the red box are of the same frequency, and are postselected for {\bf b}, which displays the distribution of the Fourier amplitudes (at the same peak frequency) on the complex plane. {\bf c} Single-shot and average transmission signals over 36 realizations with a same stable frequency indicated by the red box in {\bf a}.}
		\label{fig:FigA5}
	\end{figure*}

\noindent
	As described in the main text, the oscillating signal will synchronize to a stable frequency in each single-shot realization of the quench dynamics. However, two independent realizations separated by a very long time can still possess very different stable frequencies, due to the drift of experimental conditions, e.g., room temperatures and magnetic noises. For the measurement performed in Fig.~\ref{fig:Fig3}{\bf c}, we record 200 independent realizations of the quench dynamics, which are separated into 50 groups, with a 125~ms duration between successive realizations within each group and a 10~s duration between nearest groups. As shown in Extended Data Fig.~\ref{fig:FigA5}{\bf a}, the extracted peak frequencies of the DFT spectrum (performed for the time window 50-100~ms, already in the stable periodic region) for these 200 realizations have a long-term drift, approximately 6 times of the DFT resolution. To control the influence of frequency drifts, we select the first 100 realizations for Fig.~\ref{fig:Fig3}{\bf c}. The evolution of the peak frequency obeys the same law if we choose the full 200 realizations, albeit with a slightly larger error bar.
	
	Here, we also find that a stable oscillation pattern can have a random initial phase. To avoid the phase uncertainty caused by fluctuations of the oscillation frequency, we further postselect the data of the same frequency from the first 100 realizations (indicated by the red box in Extended Data Fig.~\ref{fig:FigA5}{\bf a}. As shown in Extended Data Fig.~\ref{fig:FigA5}{\bf b}, the phase of the Fourier amplitude $A(\omega)$ at the peak frequency $\omega_c$ is randomly distributed over $[0,2\pi)$.

	Such a random phase fluctuation has been related to a spontaneous time translation symmetry breaking in previous works. In fact, it can also be related to a phase diffusion process associated with the unbounded growth of quantum fluctuations in a time-crystalline order \cite{chan2015limit,carollo2022exactM}: while a single-run measurement always exhibits a persistent oscillation, quantum fluctuations can lead to damping of the average signal [see Extended Data Fig.~\ref{fig:FigA5}{\bf c}]. Since we cannot completely rule out classical fluctuations (e.g., random atomic motions and phase noise of the driving lasers), the relative contribution of quantum fluctuations is difficult to determine in the current experiment. In the main text, we focus on the autocorrelation function, which is also an efficient way to characterize a time crystal.\\

	\noindent\textbf{Role of the ionization process}\par\noindent
	In a hot Rydberg gas, atomic collisions can lead to strong ionization and plasma formation. The interaction between atoms and the resulting ions can even dominate over the Rydberg-Rydberg interaction in certain regimes \cite{weller2016charge,wade2018terahertz,weller2019interplay}. In this subsection, we present several experimental efforts to identify the role of ionic interactions.
	
	In Ref.~\cite{weller2016charge}, the authors ruled out Rydberg-Rydberg interactions as a possible mechanism to induce the bistability, based on the fact that the observed red energy shift of the transmission spectrum could not be explained by a repulsive vdW interaction between $|nS\rangle$ Rydberg states of Rb atoms, but instead was consistent with the attractive interactions between Rydberg atoms and ions. In our experiments, Rb atoms are excited to $ |nD_{5/2}\rangle $ Rydberg state, where the vdW interactions are attractive with a dispersive coefficient $C_6 < 0$. This attractive vdW interaction is qualitatively consistent with the red shifted transmission spectrum we observe in the experiment. Hence, the spectral shape alone is insufficient to rule out Rydberg-Rydberg atom interactions as the dominant mechanism.
	
	Second, we note that our experimental conditions are quite different from the mentioned experiments where ion-induced interaction has been identified as the dominant one. For example, we do not heat the atomic vapor and therefore operate at significantly lower densities of our room-temperature gases. Moreover, our control Rabi frequency $\Omega_c\sim1~$MHz is much lower compared to other bistability experiments, which implies a lower Rydberg excitation fraction in a hot gas.
	
	The results of these different conditions can be seen from the measured spectra. The widths of our spectra [see Figs.~\ref{fig:Fig2}{\bf b} and \ref{fig:Fig2}{\bf c}] are about an order of magnitude narrower than other bistability experiments, where the broad linewidth is attributed to the ionic interaction induced broadening.
	
\begin{figure*}[h]
		\centering
		\includegraphics[width=\linewidth]{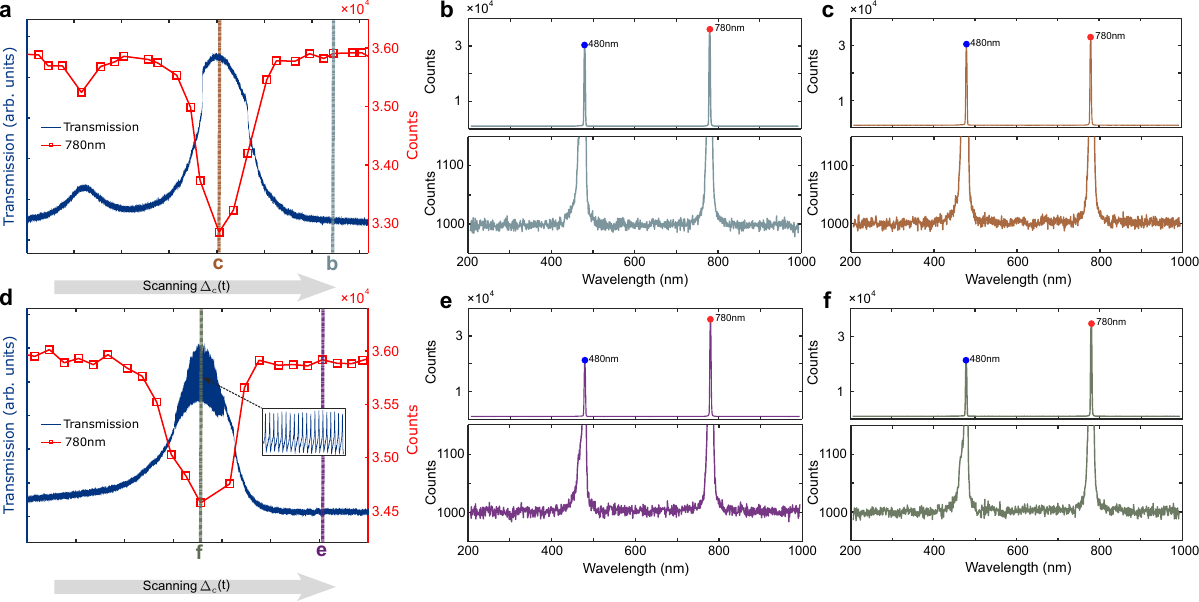}% Here is how to import EPS art
		\caption{{\bf Fluorescence measurements in our Rydberg gas. } {\bf a} and {\bf d} are transmission signals at different principal quantum number $n=55$ and $n=75$, respectively. Orange squares represent the counts of 780 nm phontons collected by the spectrometer. {\bf b}-{\bf c} and {\bf e}-{\bf f} show the full spectra from 200 nm to 1000 nm at different dutunings indicated in {\bf a} and {\bf d}.}
		\label{fig:FigA6}
	\end{figure*}

	To further search for the presence of plasma, we have performed additional measurements of the fluorescence spectrum. As discussed in Ref.~\cite{weller2019interplay}, the presence of a plasma should lead to radiative emission at a broad range of wavelength due to recombination and population of different Rydberg levels. First, we confirmed that our spectrometer reliably detects the fluorescence from the vapor cell, by monitoring the 780 nm photons when scanning the detuning. As shown in Figs.~\ref{fig:FigA6}{\bf a} and \ref{fig:FigA6}{\bf d}, the counts of 780 nm photons decrease significantly around the transmission peaks, which is a direct consequence of the reduced occupation of the intermediate state $|5P_{3/2}\rangle$ due to the electromagnetic induced transparency (EIT). Figures~\ref{fig:FigA6}{\bf a}, {\bf c}, {\bf e}, and {\bf f}, show the entire spectrum from 200 nm to 1000 nm at different detunings of the driving field labeled in {\bf a} and {\bf d}. Within the sensitivity limit of our spectrometer, we observe no pronounced fluorescence except from the two driving lasers at 780 nm (probe field) and 480 nm (coupling field), and the same background values for the on-resonant [{\bf c} and {\bf f}] and the far-detuned [{\bf b} and {\bf e}] excitation. This is consistent with fluorescence imaging of the vapor cell using a qCMOS camera (HAMAMATSU, C15550-20UP) for bandpass-filtered wavelengths between 500 nm and 700 nm after background subtraction, where no obvious fluorescence signal is detected. 
	
	Although the current experiment does not show clear evidence of ionic excitation or plasma formation, we cannot completely rule out them due to the limited resolution of the fluorescence measurement. Therefore, the ionic interaction might still be relevant to the dynamics. In fact, the inclusion of ionic interactions can also be described by the mean-field model we developed and does not destroy the mechanism towards establishment of the limit cycle oscillation. As ions are generated from Rydberg atoms, their density should be proportional to the total Rydberg density, which can give rise to a nonlinear energy shift $E_\mathrm{NL}=\chi({n}_{r}+{n}_{s})$ due to the ion-induced Stark shift \cite{weller2019interplay}, similar to Rydberg-interaction induced mean-field level shift. Thus, the nonlinear competition between distinct Rydberg states also exists when taking the ionization process into consideration, and we only need to modify the effective nonlinear coefficient $\chi$ according to the details of the ionic interaction. A more accurate and complete model considering charges needs to be investigated in the future.
	
	\noindent \textbf{Data availability}\par\noindent The data are available from the corresponding author on reasonable request.
	
	\noindent \textbf{Code availability}\par\noindent The codes are available upon reasonable request from the corresponding author.

\end{document}